\documentclass[preprint,12pt]{elsarticle}

\usepackage{graphics}

\usepackage{amssymb}
\usepackage{mathtools}
\usepackage{algorithm}
\usepackage{algpseudocode}
\usepackage{xcolor}
\usepackage{listings}
\definecolor{light-gray}{gray}{0.89}
\usepackage{blindtext}

\newcommand{\codebox}[1]{\colorbox{light-gray}{\lstinline{#1}}}

\usepackage{lineno}

\newcounter{bla}

 \journal{Computer Physics Communications}

\begin{document}

\begin{frontmatter}



\title{BetheSF: Efficient computation of the exact {tagged-particle} propagator in {single-file} systems via the Bethe eigenspectrum}


\author[a]{Alessio Lapolla\corref{author}}
\author[a]{Alja\v{z} Godec}

\cortext[author] {Corresponding author.\\\textit{E-mail address:} alessio.lapolla@mpibpc.mpg.de}
\address[a]{Mathematical bioPhysics Group, Max Planck Institute for
  Biophysical Chemistry, Am Fassberg 11, 37077 G\"{o}ttingen, Germany}

\begin{abstract}
Single-file diffusion is a paradigm for strongly correlated classical stochastic many-body
dynamics and has widespread applications in soft condensed matter and
biophysics. However, exact results for {single-file} systems are sparse and
limited to the simplest scenarios. 
We present an algorithm for computing the non-Markovian time-dependent conditional
probability density function of a {tagged-particle} in
a {single-file} of
$N$ particles diffusing in a confining external potential. The
algorithm implements an eigenexpansion of the full interacting many-body
problem obtained by means of the coordinate Bethe ansatz. While
formally exact, the Bethe eigenspectrum involves the generation and
evaluation of  permutations, {which becomes unfeasible
  for single-files with an increasing number of particles $N$.} Here we exploit the underlying {exchange}
symmetries between the particles to the left and to the right of the {tagged-particle} and show that
it is possible to reduce the complexity of the algorithm from
the worst case scenario $\mathcal{O}(N!)$ down to
$\mathcal{O}(N)$. A C++ code to calculate the non-Markovian probability density function using this algorithm is provided.
Solutions for simple model potentials are
readily implemented incl. {single-file diffusion} in a flat and a 'tilted' box, as well as in
a parabolic potential. Notably, the program allows for implementations of
solutions in arbitrary external potentials under the condition that
the user can supply solutions to the respective single-particle eigenspectra.  
\end{abstract}


\end{frontmatter}
{\bf PROGRAM SUMMARY}

\begin{small}
\noindent
{\em Program Title: BetheSF}\\
{\em Licensing provisions: MIT}\\
{\em Programming language: C++ (C++17 support required)}\\
{\em Supplementary material: makefile, README, SingleFileBluePrint.hpp}\\
{\em Nature of problem:}\\
  Diffusive {single-file}s are mathematical models of effectively
  one-dimensional strongly correlated many-body systems. While the
  dynamics of the full system is Markovian, the diffusion of a
  tracer-particle in a {single-file} is an example of non-Markovian and anomalous diffusion.
  The many-body Fokker-Planck equation governing the system's dynamics
  can be solved using the coordinate Bethe ansatz. {A na\"ive implementation
  of such a solution runs in non-polynomial time} since it requires the generation of permutations of the elements of a multiset.\\
{\em Solution method:}\\
  In this paper we show how, exploiting the exchange symmetries of the
  system, it is possible to reduce the complexity of the algorithm to
  evaluate the solution,  {using a permutation-generation algorithm,
  from $\mathcal{O}(N!)$ in the worst case scenario
    to $\mathcal{O}(N)$ in the best case scenario, which  corresponds
    to tagging the first
    or the last particle, where $N$ stands for the number
  of particles in the single-file}.\\
{\em Additional comments including Restrictions and Unusual features:}
  The code may overflow for large {single-file}s $N\geq 170$.
  All the benchmarks ran on the following CPU: Intel Xeon E3-1270 v2 3.50 GHz	4 cores. The compiler used is g++  7.3.1 (SUSE Linux) with the optimization -O3 turned on. {The code to produce all the data in the figures is included in the files: \emph{figure1.cpp}, \emph{figure2.cpp}, \emph{figure3.cpp}, \emph{figure4a.cpp} and \emph{figure4b.cpp}.}
\end{small}

\section{Introduction}
 Single-file diffusion refers to {the dynamics of one-dimensional
 systems composed of identical hard-core particles, that is, to
 many-particle diffusion subject to non-crossing boundary
 conditions. Diffusive single-file models are a
 paradigm for the
 stochastic dynamics of classical strongly correlated many-body
 systems. As such they have been studied extensively both theoretically 
 (see e.g. \cite{harris_diffusion_1965,jepsen_dynamics_1965,van_beijeren_fluctuations_1991,rodenbeck_calculating_1998,barkai_theory_2009,ambjornsson_single-file_2008,lizana_single-file_2008,lizana_foundation_2010,lapolla_unfolding_2018,lapolla_manifestations_2019})
 as well as experimentally
 (\cite{lutz_single-file_2004,lin_random_2005,
   locatelli_single-file_2016}). Single-file  diffusion underlies the
 dynamics in biological channels  \cite{hummer_water_2001}, molecular
 search processes of transcription  factors in gene regulation
 \cite{Ahlberg_2015}, transport in zeolites
 \cite{chou_entropy-driven_1999,karger_diffusion_1992} and superionic
 conductors \cite{richards_theory_1977}, and diverse phenomena in
 soft matter systems \cite{taloni_single_2017}}. 
 
 {Whereas the dynamics of the entire $N$-particle
   single-file is Markovian, the  typically observed
   ``tagged-particle'' diffusion --  the projection
   of the many-body dynamics onto the motion of a single 
 tracer particle -- is strongly
 non-Markovian \cite{lapolla_manifestations_2019}. Namely, by focusing
 on a tagged-particle alone, the $N-1$ remaining so-called latent degrees
 of freedom (i.e. the coordinates of the remaining particles) that
 become coarse-grained out, \emph{relax on exactly the same time scale} as the tagged
 particle \cite{lapolla_manifestations_2019}. This renders single-file
 diffusion somewhat special as compared to other physical examples
 probing low-dimensional projections, such as for example the dynamics
 of individual protein molecules \cite{Hu_2015} involving degrees
 freedom with relaxation times that span several orders of magnitude
 in time \cite{henzler-wildman_dynamic_2007}. As there are no ``fast''
 degrees of freedom in a single-file, low-dimensional projections give rise
 to strong memory effects, i.e. the Markov property is said to be
 strongly broken. In other words, the dynamics of a tagged-particle
 is fundamentally different (by extent as well as duration) from the
 adiabatic, Markovian approximation of the  dynamics of a single particle diffusing in a potential of mean force
 created if the remaining particles were to relax to equilibrium
 instantaneously \cite{lapolla_manifestations_2019}.}

  {Tagged-particle  diffusion in a single-file is also a representative
 toy model for diffusion in 
 so-called crowded systems, in particular when the dynamics is effectively
 one-dimensional and anomalous \cite{li_effects_2009}, 
 i.e. when the mean squared displacement of a particle $\langle
 (x(t)-x(0))^2\rangle \propto t^\alpha$ (where $\langle\cdot\rangle$
 denotes the average over an ensemble of trajectories) is not linear
 in time  as in the case of (normal)
 Brownian motion (i.e. $\alpha_{\rm Brown}=1$) but scales sub-linearly with $\alpha=1/2$,
 which is referred to as subdiffusion}
 \cite{barkai_diffusion_2010}.
 {The theoretical analysis of tagged-particle} dynamics has been carried out by several different
 techniques: the so-called ``reflection principle`` applicable to
 {single-file}s with both finite and infinite number of elements
 \cite{rodenbeck_calculating_1998}, Jepsen mapping for the central
 particle in  a finite 
 \cite{barkai_theory_2009} {or infinite single-file
   \cite{leibovich_everlasting_2013}}, the  so-called momentum Bethe
 ansatz for a finite {single-file} \cite{lizana_single-file_2008}, harmonization techniques for
 infinite {single-file}s \cite{lizana_foundation_2010}, etc.

 Here, we focus on the propagator (or the ''non-Markovian Green's
 function``) of a {tagged-particle} in a finite
 {single-file} of $N$ particles diffusing in an
 arbitrary confining potential,  that is,  the conditional probability
 density function to find the {tagged-particle} at
 position $x$ at a time $\tau$ assuming
 that at $\tau=0$ it was at $x_0$, {while the
   positions of the remaining $N-1$ particles were drawn from the
   equilibrium distribution compatible with the initial position of
   the {tagged-particle}}. 
 {In the past few years a number of detailed analyses of ensemble-
   \cite{lizana_single-file_2008,lapolla_manifestations_2019} and
   time- \cite{lapolla_unfolding_2018,lapolla_manifestations_2019}
   averaged physical observables have been carried out focusing on the motion of a tagged-particle in a single-file, which provided a generic, conceptual insight
   into the emergence of memory in projection-induced non-Markovian
   dynamics.}

 {In our previous work
 \cite{lapolla_unfolding_2018,lapolla_manifestations_2019}
 we determined} the propagator exactly by means of the
 coordinate Bethe ansatz (CBA)
 \cite{korepin_v_e_quantum_1997}. {The power of the
   CBA lies in the fact that it diagonalizes the many-body
   Fokker-Planck operator that governs the dynamics of the single-file. In other words, it expresses the dynamics of the full
   $N$-body system in a given potential in terms of a complete set of eigenfunctions and
   corresponding eigenvalues, which describe exactly how the system relaxes to
   equilibrium in terms of irreducible collective
 relaxation modes on different time-scales. By projecting these
 collective modes onto the motion of a tagged-particle we were able to
 disentangle the microscopic, collective origin of subdiffusion and
 memory in tagged-particle dynamics in simple confining potentials
 \cite{lapolla_unfolding_2018,lapolla_manifestations_2019,Lapolla_2020}.}

 {However,
 the implementation of the analytical results obtained by the
 CBA poses a computational challenge} since it
 involves an algorithm {whose complexity is non-polynomial in $N$}.
 Here we present an efficient algorithm (that in some cases runs in
 polynomial time) for evaluating the {tagged-particle}
 propagator {that exploits the exchange-symmetry of
   the problem}.
 We also present a C++ code to perform such a computation for selected
 examples. The code is easily extendable to other potentials.

 {Notably, a common alternative method to analyze tagged-particle
 dynamics in finite single-files is to perform Brownian Dynamics computer simulations.} To do
 so efficient algorithms have been designed 
 {based on the Gillespie algorithm\cite{ambjornsson_single-file_2008}, on the Ermak algorithm \cite{herrera-velarde_superparamagnetic_2008} or on the Verlet algorithm \cite{herrera-velarde_one-dimensional_2016}.}
 Nevertheless, these algorithms may still suffer from time- and space- discretization artifacts since they only
 provide an approximate solution to the problem. {Moreover, they do not
 readily reveal the collective relaxation eigenmodes, nor do they
 establish how these affect tagged-particle motion.}
 In addition, the computational cost of such Brownian Dynamics simulations is
 much larger than the one of the present
 algorithm {(for a comparison see Section \ref{comparison simulation}).}

 \section{Problem and solution by means of the coordinate Bethe ansatz}
 The {evolution} of the (Markovian) probability density function of a diffusive
 {single-file} of $N$ particles in the over-damped regime under the influence
 of an external force $F(x)=-\partial_x U(x)$,
 $G(\mathbf{x},\tau|\mathbf{x}_0)$, evolving from an initial condition
 $G(\mathbf{x},0|\mathbf{x}_0)=\delta(\mathbf{x}-\mathbf{x}_0)$ is described by the Fokker-Planck equation
 \begin{eqnarray}
  \left[\partial_\tau-\sum_{i=1}^N\left( D\partial_{x_i}^2-\mu\partial_{x_i}F(x_i)\right)\right]G(\mathbf{x},\tau|\mathbf{x}_0)&=&0,\nonumber\\
  \left(\partial_{x_{i+1}}-\partial_{x_i}\right)G(\mathbf{x},\tau,\mathbf{x}_0)\left|_{x_{i+1}=x_i}\right.&=&0,
  \quad \forall i,
  \label{FPE}
 \end{eqnarray}
 where $D$ is the diffusion coefficient, $\mu=D/k_{\mathrm{B}}T$ is
 the mobility {given by the fluctuation-dissipation theorem}, and
 $\delta(\mathbf{x}-\mathbf{x}_0)=\prod_{i=1}^N
 \delta(x_i-x_{0i})$. Eq.~(\ref{FPE}) is accompanied
 by appropriate external boundary conditions for the first and last
 particle of the {single-file}. Here we will only consider so-called
 natural ('zero probability at infinity', i.e. $\lim_{|\mathbf{x}|\to\infty}G(\mathbf{x},\tau|\mathbf{x}_0)=0$) or
 reflecting ('zero flux') boundary
 conditions, which are selected according to the specific nature of the external potential $U(x)$.
 We will assume that $U(x)$ is sufficiently confining to assure that
 the eigenspectrum of the generator $\hat{L}_N\equiv\sum_{i=1}^N[
   D\partial_{x_i}^2-\mu\partial_{x_i}F(x_i)]$ is discrete
 \cite{chupin_fokker-planck_2010}. In Eq.~(\ref{FPE}) we assumed that
 each particle experiences the same external force $F(x)$ and throughout we
 will assume that $D$ is equal for all
 particles. {Note that the corresponding over-damped
   (It\^o) Langevin equation that describes individual trajectories of the single-file and would be integrated numerically in a Brownian Dynamics simulation
 reads
\begin{equation}
   dx_i(t)=\mu
   F(x_i(t))dt+\sqrt{2D}dW^i_t,\,\langle dW_t\rangle=0,\, \langle
   dW^i_tdW^j_{t'}\rangle=\delta_{ij}\delta(t-t')dt,\forall i,
\label{Ito}
\end{equation}
   where $dW_t$ is an increment of the
   Wiener process (Gaussian white noise), whereby we must enforce
   that particles remain
   ordered at all times, i.e. $x_i(t)\le x_{i+1}(t),\forall i,t$.}
 
 The boundary value problem in Eq.~(\ref{FPE}) can be solved exactly by means
 of the coordinate Bethe ansatz \cite{korepin_v_e_quantum_1997}, which
 requires that we {(only)} know the eigenexpansion of the single-particle
 Green's function. That is, we are required to solve the following
 single-particle Fokker-Planck equation with the same external boundary conditions
 \begin{equation}
 (\partial_\tau -\hat{L}_1)\Gamma(x_i,\tau|x_{0i})=0
 \end{equation}
 with initial condition $\Gamma(x_i,0|x_{0i})=\delta(x_i-x_{0i})$,
 which can be conveniently expressed by means of a (bi)spectral
 expansion  
 \begin{equation}
  \Gamma(x_i,\tau|x_{0i})=\sum_{k=0}^\infty \psi_{k_i}^R(x_i)\psi_{k_i}^L(x_{0i}) \mathrm{e}^{-\lambda_{k_i} \tau},
 \end{equation}
 where $-\lambda_{k_i}<0, \forall i>0$ and $\lambda_0=0$ are the eigenvalues, and $\psi_{k_i}^{L/R}(x)$ are respectively the $k_i$th left and the right
 eigenfunction of the operator $\hat{L}_1$, which form a complete
 bi-orthonormal basis. {Here we assume
 detailed balance to be obeyed and hence}
 $\psi_l^R(x)\propto \mathrm{e}^{-\beta U(x)}\psi_l^L(x)$
 \cite{kurchan_six_2009}, where $\beta=1/(k_B T)$ is the inverse of
 the thermal energy. The solution to the many-body Fokker-Planck
 equation can be written as
 \begin{equation}
  G(\mathbf{x},\tau|\mathbf{x}_0)=\sum_{\mathbf{k}} \Psi_\mathbf{k}^R(\mathbf{x})\Psi_\mathbf{k}^L(\mathbf{x}_0)\mathrm{e}^{-\Lambda_\mathbf{k}\tau}.
    \label{green}
 \end{equation}
The many-body eigenvalue $\mathbf{k}$ corresponds to a multiset
containing the $N$ natural numbers $\{k_1,k_2,\cdots,k_N\}$ and
$\mathbf{0}$ denotes the unique ground state of the many-body system
in which each single-particle eigenvalue is equal to zero.  
{Each pair of many-body eigenvalues and eigenfunctions satisfies the eigenvalue problem}
 \begin{equation}
  \hat{L}_N\Psi_\mathbf{k}^R=\Lambda_\mathbf{k}\Psi_\mathbf{k}^R.
 \end{equation}
 The Bethe ansatz solution postulates that the right eigenfunction has the following form
 \begin{equation}
  \Psi_\mathbf{k}^R = \hat{O}_\mathbf{x}\sum_{\{\mathbf{k}\}} \prod_{i=1}^N c_i\psi_{k_i}^R(x_i);
 \end{equation}
  where $\sum_{\{\mathbf{k}\}}$ denotes the sum over all the possible
  permutations of the multiset $\mathbf{k}$ (see \ref{permutations})
  and $\hat{O}_\mathbf{x}$ denotes the particle-ordering operator
  defined as
 \begin{equation}
   \hat{O}_\mathbf{x}\equiv \prod_{i=2}^N \Theta(x_{i}-x_{i-1}),
 \end{equation}
 where $\Theta(x)$ denotes the Heaviside step function.
 
 The $N$ constants $\{c_i\}$ and the many-body eigenvalue are fixed
 imposing the $N-1$ internal boundary conditions in Eq.~(\ref{FPE})
 alongside the pair of external boundary conditions. This leads to the many-body eigenvalue
 \begin{equation}
   \Lambda_\mathbf{k}=\sum_{i=1}^N \lambda_{k_i},
 \label{EV}  
 \end{equation}
 and in the case of zero-flux boundary conditions all $c_i$ turn out
 to be equal to one. Finally, a proper orthonormalization between left and right many-body eigenfunctions must be assured, for example
 \begin{equation}
  \Psi_\mathbf{k}^{L/R}(\mathbf{x})= \mathcal{N}^{-1/2}
  \hat{O}_\mathbf{x}\sum_{\{\mathbf{k}\}} \prod_{i=1}^N
  \psi_{k_i}^{L/R}(x_i),
 \label{EBethe} 
 \end{equation}
 where the normalization factor $\mathcal{N}$ is equal to the number
 of permutations of the multiset $\mathbf{k}$ (see
 \ref{permutations}).
 
 Here we are interested in the non-Markovian Green's function referring
 to the propagation of a {tagged-particle} starting from a fixed initial
 condition $x_{0i}$ while the remaining particles are drawn from those
 equilibrium configurations that are compatible with the initial
 condition of the {tagged-particle} \cite{lapolla_manifestations_2019}
 \begin{equation}
  \mathcal{G}(x_i,\tau|x_{0i})=V_{\mathbf{0}\mathbf{0}}^{-1}(x_{0i})\sum_\mathbf{k} V_{\mathbf{0}\mathbf{k}}(x_i)V_{\mathbf{k}\mathbf{0}}(x_{0i})\mathrm{e}^{-\Lambda_\mathbf{k}\tau},
\label{NMGreen}
 \end{equation}
 where the 'overlap elements' are defined as 
 \begin{equation}
  V_{\mathbf{k}\mathbf{l}}(z)=\int d\mathbf{x} \delta(z-x_i) \Psi_\mathbf{k}^L(\mathbf{x}) \Psi_\mathbf{l}^R(\mathbf{x}),
  \label{overlap integral}
 \end{equation}
 and $\delta(x)$ is Dirac's delta. In the specific case of
 equilibrated initial conditions for background particles only the special cases 
 \begin{eqnarray}
  V_{\mathbf{k}\mathbf{0}}(z)&=&\int d\mathbf{x} \delta(z-x_i) \Psi_\mathbf{k}^L(\mathbf{x}) \Psi_\mathbf{0}^R(\mathbf{x}),\nonumber\\
  V_{\mathbf{0}\mathbf{k}}(z)&=&\int d\mathbf{x} \delta(z-x_i)
  \Psi_\mathbf{0}^L(\mathbf{x}) \Psi_\mathbf{k}^R(\mathbf{x})
 \label{integrals} 
 \end{eqnarray}
 are important. {Note that any numerical
   implementation of Eq.~(\ref{NMGreen}) involves a truncation at some
 maximal eigenvalue $\Lambda_\mathbf{M}$.}
 The ordering operator allows us to evaluate the integrals
 (\ref{integrals}) as nested integrals, i.e.
 \begin{equation}
 \mathop{\mathrlap{\int}{\,}n}_a^b f(\mathbf{x}) d\mathbf{x}=
 \int_a^b dx_{1}\int_{x_1}^{b} dx_{2}\cdots\int_{x_{N-2}}^{b}dx_{N-1}\int_{x_{N-1}}^{b} dx_{N}f(\mathbf{x}).
 \end{equation}
 Since by construction the integrand is invariant under exchange of
 the $\{x_i\}$ coordinates we can take advantage of the
 {so-called extended phase-space integration} \cite{lizana_diffusion_2009} and greatly simplify the multi-dimensional nested integral to a product of one-dimensional integrals
 \begin{equation}
  \mathop{\mathrlap{\int}{\,}n}_a^b f(\mathbf{x}) \delta(z-x_i) d\mathbf{x}=\left(\prod_{j=1}^{i-1}\int_a^z dx_j\right)\left(\prod_{j=i+1}^{N}\int_z^b dx_j\right)\frac{f(x_i=z,\{x_j, j\neq i\}) }{N_L!N_R!},
 \end{equation}
 where $a$ and $b$ are the lower and upper boundary of the domain, respectively, and $N_L$($N_R$) is the number of particles to the left(right) of the tagged one.
 These last two equations allow us the write Eq.~(\ref{overlap integral}) as
 \begin{equation}
  V_{\mathbf{k}\mathbf{l}}(z)=\frac{m_\mathbf{l}}{N_L!N_R!}\sum_{\{\mathbf{k}\}}\sum_{\{\mathbf{l}\}}T_i(z)\prod_{j=1}^{i-1} L_j(z)\prod_{j=i+1}^N R_j(z),
  \label{overlap}
 \end{equation}
 where $m_\mathbf{l}$ is the multiplicity of the multiset $\mathbf{l}$
 defined in \ref{permutations} and we have introduced the auxiliary functions
 \begin{subequations}
 \begin{equation}
  T_i(z)=\psi^L_{k_i}(z)\psi^R_{l_i}(z),
 \end{equation}
 \begin{equation}
  L_j(z)=\int_a^z dx \psi^L_{k_j}(x)\psi^R_{l_j}(x),
 \end{equation}
 \begin{equation}
  R_j(z)=\int_z^b dx \psi^L_{k_j}(x)\psi^R_{l_j}(x).
 \end{equation}
 \label{potential equations}
 \end{subequations}
 {Once substituted into Eq.~(\ref{NMGreen})
  Eqs.~(\ref{overlap}-\ref{potential equations}) deliver the tagged
  particle propagator sought for.}

 \section{Avoiding permutations}
  \begin{algorithm}
  \caption{Calculate $V_{\mathbf{k}\mathbf{l}}(z)$}
  \begin{algorithmic}[1]
  \Require \textcolor{white}{.} \begin{itemize}
            \item $\mathbf{k},\mathbf{l}$ multisets;
            \item $z\in\mathbb{R}, a\leq z\leq b$;
            \item functions: $T_i(z), L_j(z), R_j(z)$;
            \item a function to generate all the permutation of multiset $P(\mathbf{k})$;
            \item a function to calculate the number of permutation of a multiset: $\mathcal{N}_\mathbf{k}$;
            \item a function to generate all the $t$-combinations of a multiset $C(\mathbf{k},t)$;
            \item a function to compute the multiset difference $\mathbf{k}\setminus\mathbf{l}$;
            \item a function to create the largest set from a multiset $\tilde{k}=S(\mathbf{k})$;
   \end{itemize}
 
  \State calculate $\mathcal{N}_\mathbf{k}$ and $\mathcal{N}_\mathbf{l}$ and pick the multiset with the smallest number of permutations (let us assume it is $\mathbf{l}$);
  \State initialize $s \gets 0$;
  \ForAll {$\mathbf{l}^* \in P(\mathbf{l})$}
   \State create the multiset of pairs $\mathbf{p}=\{\{k_1,l^*_1\},\cdots,\{k_N,l^*_N\}\}$;
   \State $\tilde{u}\gets S(\mathbf{p})$;
   \For {$u \in \tilde{u}$}
    \State $\mathbf{r}\gets\mathbf{p}\setminus u$;
    \State $t\gets \mathrm{min}(N_L, N_R)$;
    \State initialize $s_1\gets0$;
    \ForAll {$\mathbf{s} \in C(\mathbf{r},t)$}
     \State $\mathbf{d}\gets\mathbf{r}\setminus\mathbf{s}$
     \If {$t=N_L$}
      \State $a \gets T_i(z)\prod_{j\in \mathbf{s}} L_j(z)\prod_{j\in \mathbf{d}} R_j(z)$;
     \Else
      \State $a \gets T_i(z)\prod_{j\in \mathbf{d}} L_j(z)\prod_{j\in \mathbf{s}} R_j(z)$
     \EndIf
     \State $s_1+=\mathcal{N}_\mathbf{s}\mathcal{N}_\mathbf{d} a$;
    \EndFor
    \State $s+=s_1$;
  \EndFor
  \EndFor\\
  \Return $\frac{m_\mathbf{l}}{N_L!N_R!}s$.
  \end{algorithmic}
  \label{main algorithm}
  \end{algorithm}
  
 Although the {extended phase-space integration (cf
   Eqs.~(\ref{overlap integral}) and ~(\ref{overlap}))}  substantially
 simplifies the integrals involved in the computation of the tagged
 particle propagator we still need to sum over all the permutations of
 $\mathbf{l}$ and $\mathbf{k}$ {in Eq.~(\ref{overlap})}. A brute force (or na\"{i}ve) approach
 is thus not feasible, not even for rather small
 {single-files} since we
 need to evaluate the products
 $V_{\mathbf{0}\mathbf{k}}(x_i)V_{\mathbf{k}\mathbf{0}}(x_{0i})$ in
 Eq.~(\ref{NMGreen}) up to {$2\times N!$ times in the worst case scenario for a calculation involving only the Green's function; and for a general element $V_{\mathbf{l}\mathbf{k}}$ up to $(N!)^2$.} 
 
 The main contribution of this paper {is Algorithm \ref{main algorithm}}
 that reduces the number of terms in the Bethe ansatz solution
 {entering Eq.~(\ref{overlap})} that need
 to be computed explicitly. Namely, 
since the {full single-file} diffusion model
{is symmetric with respect to the exchange of particles} many terms
arising from the permutations of {the eigennumbers of the multisets in Eq.~(\ref{overlap})  happen to be
identical. Algorithm \ref{main algorithm} counts how many terms are equal and
computes only those that are unique, and does so only once. These
unique terms are then multiplied by their respective
multiplicity and summed up to yield the result Eq.~(\ref{NMGreen})}.
{Algorithm \ref{main algorithm} thereby avoids going
  through the large number of equivalent permutations of the multisets
 in the sum with the larger number of terms between $\sum_{\{\mathbf{k}\}}$ and $\sum_{\{\mathbf{l}\}}$ in Eq.~(\ref{overlap}).}
 In the specific case of the {tagged-particle} Green's
function defined in Eq.~(\ref{NMGreen}), where one of the two
multisets {$\{\mathbf{k}\},\{\mathbf{l}\}$}
corresponds to the ground state (having only one permutation), the
algorithm in fact avoids permutations entirely.

More {precisely} (i.e. for a general $V_{\mathbf{k}\mathbf{l}}(x_i)$), the algorithm first generates all permutations of the
multiset having the smallest number of permutations  $P(\mathbf{l})$ (for sake of simplicity let us assume that this is the multiset
$\mathbf{l}$  with $\mathcal{N}_\mathbf{l}$ distinct
permutations). Then, for each of these permutations a multiset of pairs
is created: $\mathbf{p}=\{\{k_1,l^*_1\},\cdots,\{k_N,l^*_N\}\}$.
The function $S(\mathbf{p})$ selects the largest possible set from
$\mathbf{p}$ and generates for each element $u$ of the resulting set
the 'difference multiset': $\mathbf{r}=\mathbf{p}\setminus u$. In the
following it determines $t=\mathrm{min}(N_L, N_R)$ and all the
$t-$combinations of $\mathbf{r}$ are generated via $C(\mathbf{r},t)$
(note that $t$ here does not refer to time). For each of these
combinations $\mathbf{s}$ the complementary multiset
$\mathbf{d}=\mathbf{r}\setminus\mathbf{s}$ is created and the number
of permutations of $\mathbf{s}$ and $\mathbf{d}$ is computed.
Finally, the products in Eq.~(\ref{overlap}) are calculated (where $u$ is
the pair of eigennumbers belonging to the {tagged-particle}) and
accounted for their multiplicity.

\begin{figure}
   \includegraphics[width=0.98\textwidth]{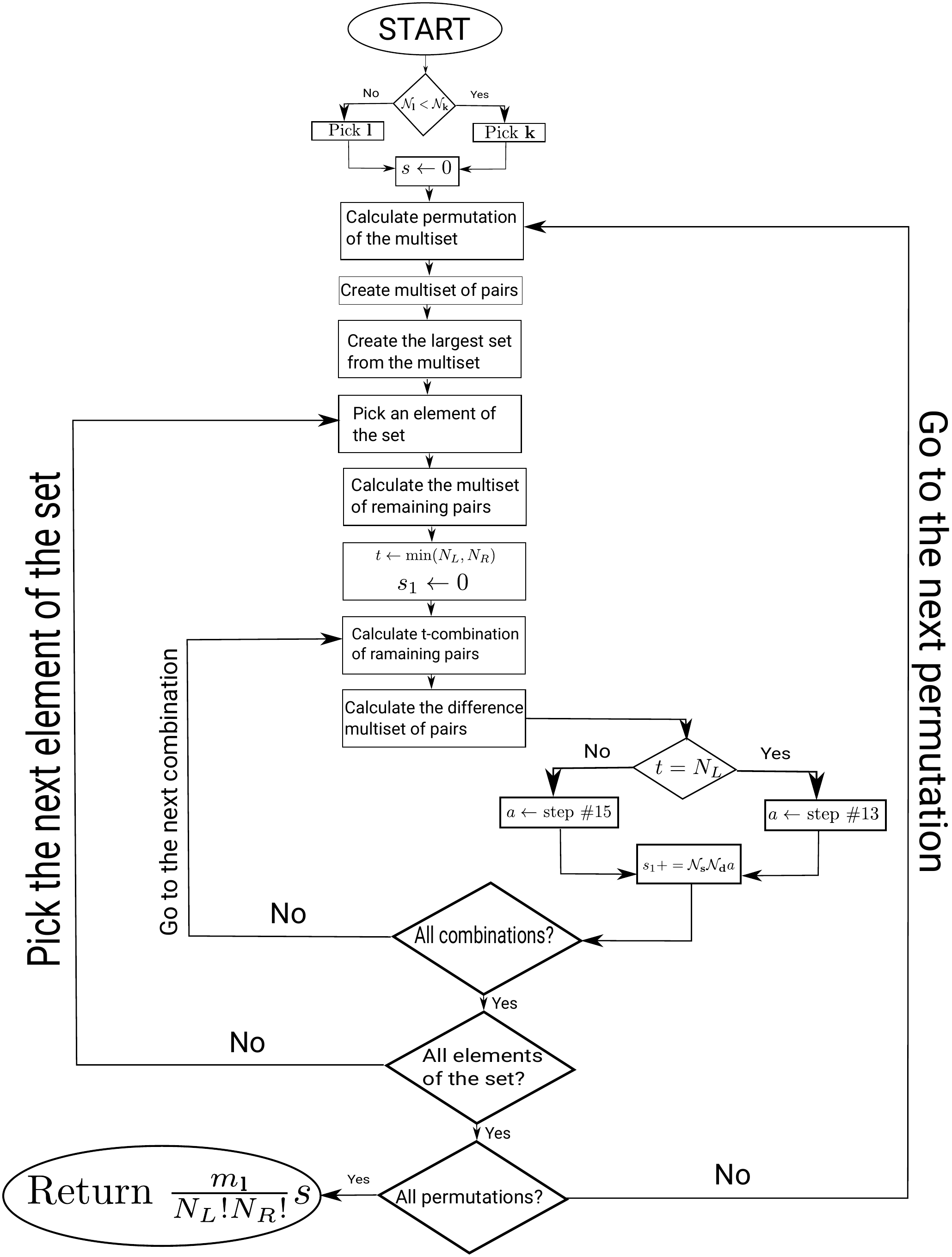}
   \caption{The flowchart of Algorithm \ref{main algorithm}. The steps
     \#13 and \#15 are not reported for spatial constraints and can be found in the explanation of the algorithm.} 
   \label{flowchart}
 \end{figure}

\begin{figure}
   \includegraphics[width=0.98\textwidth]{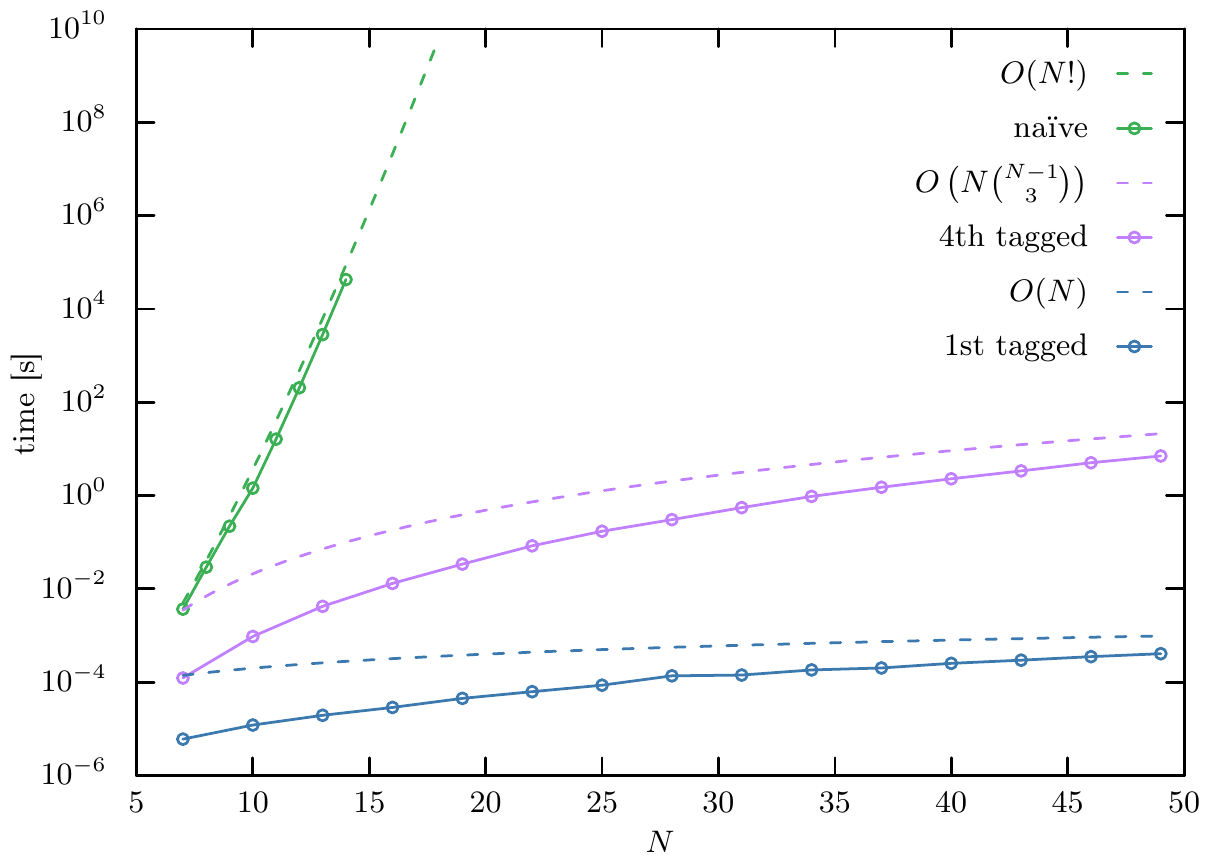}
   \caption{Computational time (in seconds) required to calculate
     $V_{\mathbf{k}\mathbf{0}}$ for a {single-file} confined to a flat
     box depending on the number of particles in the {single-file} $N$; the $\mathbf{k}$-multiset \emph{has been chosen to represent
       the worst case scenario},
     i.e. $\mathbf{k}=\{1,2,\cdots,N\}$. The green line corresponds to
     the running time of the na\"{i}ve implementation for comparison
     (which does not depend on which particle is tagged),
     while the blue and purple lines depict the running time of our
     program tagging the first and fourth particle, respectively. Dashed lines depict {the computational complexity} for the various cases.} 
   \label{algorithmiccomplexity}
 \end{figure}

{
In summary, our algorithm exploits
the fact that the extended phase-space integration allows us to
ignore the ordering of the particles to the left and to the right of
the tagged-particle, respectively. A consequence of this symmetry is
that several terms that appear in Eq.~(\ref{overlap}) are identical.
Therefore, we can substitute the permutations of one multiset in
Eq.~(\ref{overlap}) with all its combinations that are not
tied to any ordering by definition. This makes the algorithm more efficient.
}
 A pseudocode-implementation is presented in Algorithm \ref{main
   algorithm} {and an explicit flowchart is depicted in Fig.~\ref{flowchart}}. The reduction of the computational time achieved by
 our algorithm compared to a na\"{i}ve implementation is
 presented in Fig.~\ref{algorithmiccomplexity}.
 
\paragraph{Computational complexity of Algorithm \ref{main algorithm}}
 The computational complexity of the algorithm can be derived by following
 its flow {(see Fig.~\ref{flowchart})}. For the sake of simplicity we will (only initially) assume that
 the multiset $\mathbf{l}$ has only one possible permutation.  Let $\mathcal{U}$
 be the number of unique elements belonging to the multiset
 $\mathbf{k}$. Then for each unique element $u \in \mathbf{k}$ we need
 to iterate over all the $t$-combinations of the multiset
 $\mathbf{k}\setminus u$, where $t=\mathrm{min}(N_L, N_R)$. The number
 of these combinations is given by the function $\mathcal{M}(N-1,t)$
 -- an algorithm describing and computing this function is presented in
 \ref{permutations}.  Hence, the complexity of the algorithm is given
 by $\mathcal{O}(\mathcal{U}\cdot \mathcal{M}(N-1,\mathrm{min}(N_L, N_R))$. In the worst-case scenario, in which all the elements of $\mathbf{k}$ are
 different, the complexity is $\mathcal{O}\left(N\cdot
 \binom{N-1}{\mathrm{min}(N_L, N_R)}\right)$. However, even in this
 worst case scenario the algorithm scales linearly $\mathcal{O}(N)$ in the
 number of particles if we tag the first or the last particle (see
 Fig.~\ref{algorithmiccomplexity}). In the general case when $\mathbf{l}\neq
 \mathbf{0}$, i.e. the one in which $\mathbf{l}$ admits more
 permutations (which, however, is not required for evaluating Eq.~(\ref{NMGreen})), the complexity deteriorates fast since the evaluation of all permutations of $\mathbf{l}$ must be considered; {the computational complexity in this case is $\mathcal{O}(\mathcal{N}_{\mathbf{l}}\cdot \mathcal{U}\cdot \mathcal{M}(N-1,\mathrm{min}(N_L, N_R))$, where $\mathcal{N}_{\mathbf{l}}$ is the number of permutations with repetitions of $\mathbf{l}$ and we assume that $\mathcal{N}_{\mathbf{l}}\leq\mathcal{N}_{\mathbf{k}}$.}
 
\section{Implementation}
 The main goal of the code attached to this article is to compute the
 Green's function of any {tagged-particle} in a {single-file} of $N$
 elements given a potential $U(x)$. For this reason we opt for an
 \emph{object-oriented approach} that allows the user to easily extend the
 code to incorporate any potential satisfying the constraints on
 $\hat{L}_N$. The code defines the abstract base class: \codebox{class
   SingleFile} {(in \emph{SingleFile.hpp})} responsible for the interface and for the functions
 that are responsible for the computation of the overlap elements
 (Eq.~(\ref{overlap})). Conversely, all functions directly related to
 some specific potential $U(x)$ are private pure abstract base
 functions and must be implemented by the user in a derived class.
 
 In our codebase we provide three different derived classes:\\
 \codebox{class SingleFileFlat : public SingleFile;},\\
 \codebox{class SingleFileOnSlope : public SingleFile;},\\
 \codebox{class SingleFileHarmonic : public SingleFile;}
 {in the header file \emph{SingleFileDerived.hpp},} covering several different 'canonical' cases of {single-file} systems.

 \paragraph{The base class}
 The base class \codebox{class SingleFile} provides a common interface
 to all {single-file} systems.  It contains the following functions: the equilibrium
 probability density function
 \codebox{virtual double eq_prob(const
   double x) const;},\\
 the two-point joint density\\
 \codebox{double joint2dens(const double  x, const double t, const double
   x0);}\\
 and the Green's function\\
 \codebox{double green_function(const double x, const double t, const double
   x0);}\\
 for a specific {tagged-particle}
 {implementing the analytical solution in Eq.~(\ref{NMGreen})}. The function evaluating
 the equilibrium probability density function of a {tagged-particle},
 i.e. $ \mathcal{G}_{\mathrm{eq}}(x_i)=\lim_{t\to\infty} \mathcal{G}(x_i,\tau|x_{0i})$, is virtual since
 for a given potential $U(x)$ it often has a relatively
 simple form.  In addition, na\"{i}ve implementations directly computing all
 permutations have also been defined in the interface:
 \codebox{double joint2dens_naive(const double  x, const double t, const double x0);}\\
 and 
 \codebox{double green_function_naive(const double x, const double t, const double x0);}.\\
 {These two functions call the function}\\
 \codebox{ double Vkl_element_naive(std::vector<int>& k_vec, std::vector<int>& l_vec, const double x) const;}\\
 {that implements slavishly Eq.~(\ref{overlap integral})}. Finally, the
 interface of the class is completed by several tiny functions that
 allow changing the parameters of an instance of the class, like
 the {tagged-particle} or the diffusion coefficient
 $D$.
 
 The base class is also responsible for the internal machinery to use our fast
 algorithm implementing (among its private members):\\
 \codebox{double
   Vkl_element(std::vector<int>& k_vec, std::vector<int>& l_vec, const
   double x) const;};\\
 and its specialized versions, defined by default in its terms:\\
 \codebox{virtual double  V0k_element(std::vector<int>& k_vec, const double x)
   const;}
 and\\
 \codebox{virtual double Vk0_element(std::vector<int>& k_vec, const double x)
   const;}.
 
 These specialized versions are made virtual to allow a
 derived class to override them if they {require special settings} (one such
 example is the {single-file} in a linear potential). {The declarations and definitions of these functions can be found in the files \emph{SingleFile.hpp} and \emph{SingleFile.cpp}.} Our
 algorithm computes the $t$-combinations of a multiset and this
 feature is provided by the friend class
 \codebox{template<typename T>
   class UCombinations;}
that implements {(in the file \emph{combinations.hpp})} a classical
algorithm given in
\cite{knuth_donald_ervin_art_2013}. {We use
  \codebox{std::next_permutation} for the computation of permutations.}
Finally, this base abstract class
 defines the private members responsible for the calculation of the
 single-particle eigenvalues and for the evaluation of Eqs.~(\ref{potential
   equations}).
 These are pure virtual functions since they depend on
 the specific external potential, and hence they must be implemented
 by the derived class. Finally, the pure virtual function 
 \codebox{virtual int eigenfunction_condition(const int i)
   const=0;}\\
 defines the rule to initialize the private
 member\\
 \codebox{std::vector<std::vector<int>>
   eigenfunction_store;}
 that contains (row-wise) all the multisets considered in the evaluation of Eq.~(\ref{green})
 for a given specific potential $U(x)$. 
 For this reason the derived class is responsible for initializing this last
 member (in its constructor, for example). We provide the protected
 function \codebox{void eigenfunction_store_init();} to initialize this data structure (details are given below). However,
 the user may implement a different way to initialize the container as
 well, for example by importing it from an existing file. 
 
 \paragraph{The derived classes}
 \label{derived}
 Three types of analytically solvable potentials:
 $U(x)=0$, $U(x)=gx$ and $U(x)=\gamma x^2/2$,
 are implemented ($g,\gamma$ being real
  and positive).
These implementations assume that the
  positions of non-{tagged-particle}s are drawn from their respective
  equilibrium distributions conditioned on the initial position of the {tagged-particle}. The many-body Bethe eigenvalues for these models are given by
  \begin{eqnarray}
  \Lambda_\mathbf{k}&=&\sum_{i=1}^N D\pi^2 k_i^2, \label{flat eig}\\
  \Lambda_\mathbf{k}&=&\sum_{i=1}^N (1-\delta_{k_i,0})\left(D\pi^2k_i^2+\frac{g^2}{4D}\right),\label{slope eig}\\
  \Lambda_\mathbf{k}&=&\sum_{i=1}^N \gamma k_i \label{oh eig},
  \end{eqnarray}
 for  $U(x)=0$, $U(x)=gx$ and $U(x)=\gamma x^2/2$ respectively.
 {In the files \emph{SingleFileDerived.hpp} and
   \emph{SingleFileDerived.cpp} the functions related to
   single-particle solutions (further details are given in \ref{single
     eigenexpansion}) that enter the Bethe-ansatz solution in Eq.~(\ref{overlap}) are implemented as overridden private member functions of the derived classes. The function \\
 \codebox{double lambda_single(const int n) const;}
 calculates the single-particle eigenvalue while the functions\\
 \codebox{double tagged(const int lambda_k, const int lambda_l, const double x) const;}\\
 \codebox{double lefttagged(const int lambda_k, const int lambda_l, const double x) const override;}\\	
 \codebox{double righttagged(const int lambda_k, const int lambda_l, const double x) const override;}\\
 implement respectively $T_i$, $L_i$ and $R_i$ defined in
 Eq.~(\ref{potential equations}). These last four functions must be
 implemented following the template in \emph{SingleFileBluePrint.hpp}
 if the user wishes to implement a solution for a different potential $U(x)$.}
 In our implementation the constructor of a derived class takes a parameter
 \codebox{int max_many_eig}$\equiv M$. {This
   positive parameter is proportional to the maximum eigenvalues we
   want to consider in the implementation of Eq.~(\ref{NMGreen}).}
 
  {Note that by fixing the largest eigenvalue $\Lambda_\mathbf{M}$ we
    consider in the computation of the Green's function in
    Eq.~(\ref{NMGreen}) we implicitly determine the shortest
    time-scale for which the solution is reliable, i.e. the solution is exact for times $\tau\gtrsim 1/\Lambda_\mathbf{M}$ \cite{gardiner_c.w._handbook_1985, risken_fokker-planck_1996}.
 Since the eigenspectra of {single-file} systems are
 always degenerate, once fixed $M$ is used to select the multisets $\mathbf{k}$ (each of them uniquely identifies an eigenfunction) that must be considered in Eq.~(\ref{NMGreen}).
 However, the rule for selecting allowed multisets is system dependent; in the case of the flat and linear potentials (cf. Eqs.~(\ref{flat eig}) and (\ref{slope eig})) we can only
 accept multisets satisfying $\sum_{i=1}^N k_i^2\leq M$, and
 for the harmonic potential (Eq.~(\ref{oh eig})) only those satisfying $\sum_{i=1}^N k_i\leq M$ are allowed.}
 These constraints must be implemented
 in the pure abstract function\\
 \codebox{virtual int eigenfunction_condition(const int i) const=0;}.

 According to this function the constructors of our derived
 classes fill\\
 \codebox{std::vector<std::vector<int>> eigenfunction_store;}
 using a slightly modified implementation of a classical algorithm for
 computing integer partitions found in
 \cite{knuth_donald_ervin_art_2013}, which takes into account the
 possibility that one (or more) of the $k_i$ can be equal to $0$.
 This {implementation} is provided in the friend class \codebox{class
   IntegerPartitions;} {in the file \emph{IntegerPartitions.hpp}}. The number of integer partitions {(see \ref{permutations} for an example)} generated by
 this algorithm is the sum of all the possible bounded compositions of
 $N$ numbers such that their sum is between $0$ and $M$. The
 number of bounded composition of $N$ elements summing to $M$
 {(see e.g. Eq.~(\ref{EV}) alongside the specific values of
   $\lambda_{k_i}$ given in Appendix \ref{single eigenexpansion})} is
 equivalent to the $N$-combinations of multiset in which all the
 numbers between $0$ and $M$ appear at most $N$ times
 \cite{knuth_donald_ervin_art_2013}.  The function
 \codebox{virtual
   int eigenfunction_condition(const int i) const=0;} 
 then selects from
 those only the allowed ones. All these steps are wrapped in the
 aforementioned \codebox{void eigenfunction_store_init();} function. 
 { The function \codebox{virtual
   int eigenfunction_condition(const int i) const=0;} must be implemented by the user in a new derived class implementing a different potential.}\\
 
 \begin{figure}
   \includegraphics[width=0.98\textwidth]{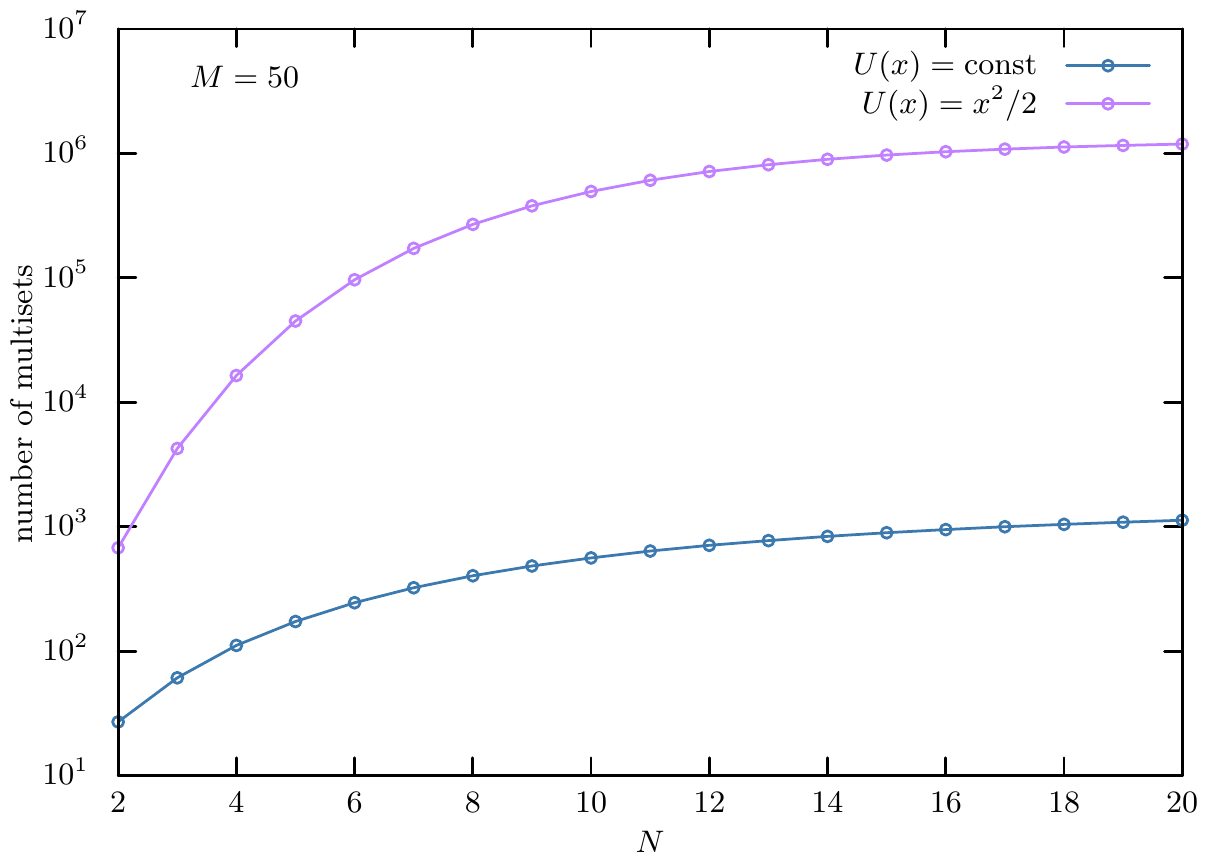}
   \caption{The number of multisets that must be considered assuming
     \codebox{max_many_eig}$\equiv M=50$ for the {single-file}
     in a flat (blue line) or in a harmonic (purple line) potential.}
  \label{number multiset}
 \end{figure}
 
 In Fig.~\ref{number multiset} we show how many multisets must be
 considered for the convergence of the sum on a time-scale
 $\tau\gtrsim1/\Lambda_\mathbf{M}$. Since these multisets are saved in 
 \codebox{std::vector<std::vector<int>> eigenfunction_store;}, 
 the size of this data structure prescribes the memory requirements of
 our program. {The program saves these values to allow for a flexible
 way to compute the non-Markovian Green's function (\ref{NMGreen}) for the same system
 when tagging a different particle without the need to re-compute the
 necessary multisets.} Though this number can become huge for some
 systems, for example in the case of the harmonic potential, it has
 been proved that for regular Sturm-Liouville problems the eigenvalues
 scale quadratically for large $k_i$
 \cite{hochstadt_asymptotic_1961}. Since often also non-regular
 Sturm-Liouville problems on a infinite domain are treated numerically
 using truncation methods \cite{boyd_chebyshev_2001} our choice to
 save these numbers to enhance the flexibility and readability of the
 code is justified. Nevertheless, it would be equally possible not to
 save the necessary multisets and instead do all calculations on-the-fly.

 Moreover, {according to the specific properties of
   the Fokker-Planck operator $\hat{L}_N$ we sometimes find that
 $V_{\mathbf{k}\mathbf{0}}(x)=V_{\mathbf{0}\mathbf{k}}(x)$}. A a result
 both functions can be implemented in terms of a single
 function responsible for $V_{\mathbf{k}\mathbf{l}}(x)$ without the
 necessity of code duplication. However, for the {single-file} in a
 linear potential this is not the case
 \cite{lapolla_manifestations_2019}. For this reason the functions for
 calculating the 'overlaps' (i.e. Eqs.~(\ref{overlap integral})) with the ground state are virtual, such
that  they can be implemented without re-factoring the code. In our
 implementation of \codebox{class SingleFileOnSlope;} the function
 \codebox{Vk0_element} is overridden with a marginally faster
 version to take into account the asymmetry {of the single-file in a
 linear potential}.
 
 \begin{figure}
   \includegraphics[width=0.98\textwidth]{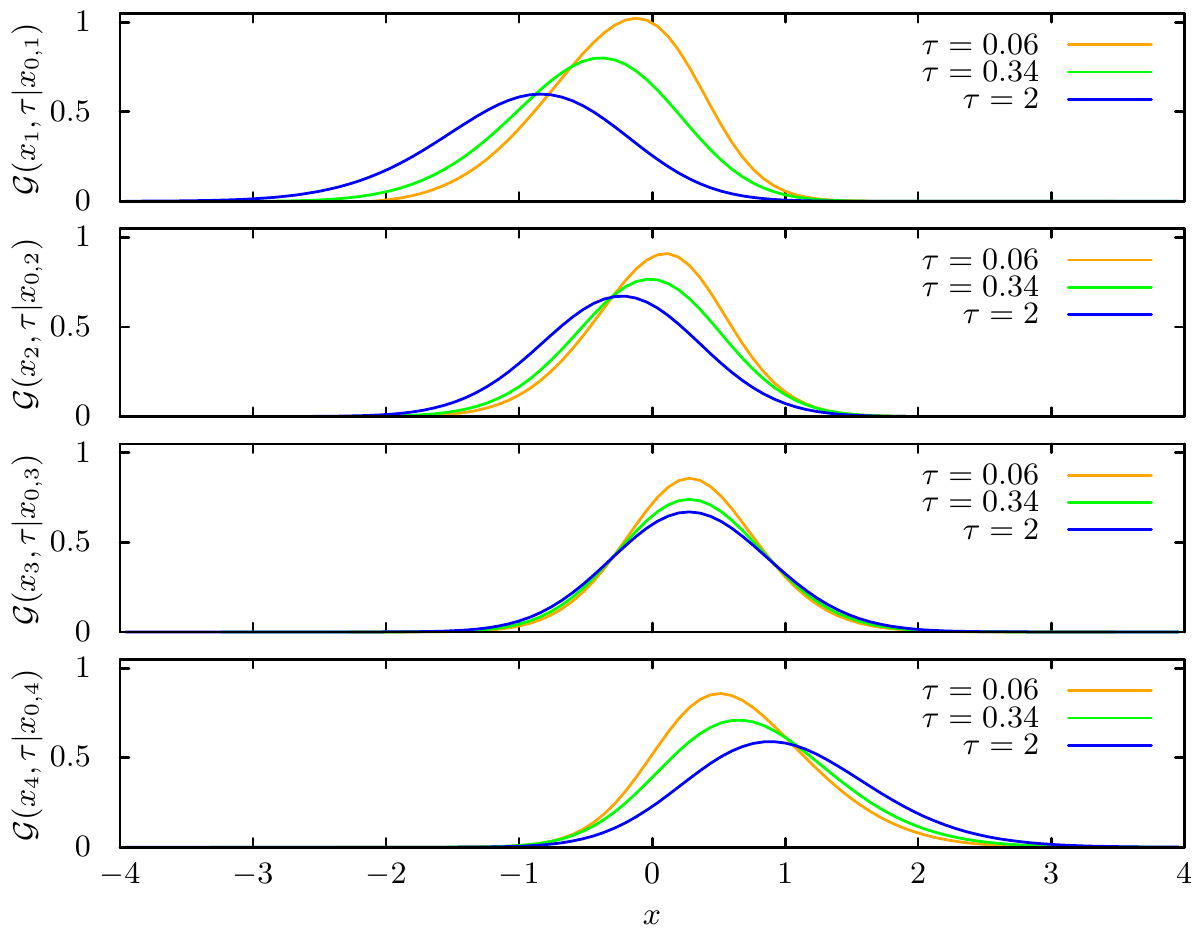}
    \caption{The {tagged-particle} Green's function for different particles of a {single-file} of $4$ particles in a harmonic potential $U(x)=x^2/2$ at different times (the relaxation time is $\sim\lambda_1^{-1}=1$). Each particle's initial position is $x_0=0.305$ assuming that all the other particles are in their respective equilibrium conditioned on the position of the {tagged-particle}.}
   \label{green_harmonic}
 \end{figure}
 In order to illustrate our final result we depict  in Fig. \ref{green_harmonic} the computed Green's function for a {single-file} of $4$ particles in a harmonic potential $U(x)=x^2/2$.
 \begin{figure}
    \includegraphics[width=0.98\textwidth]{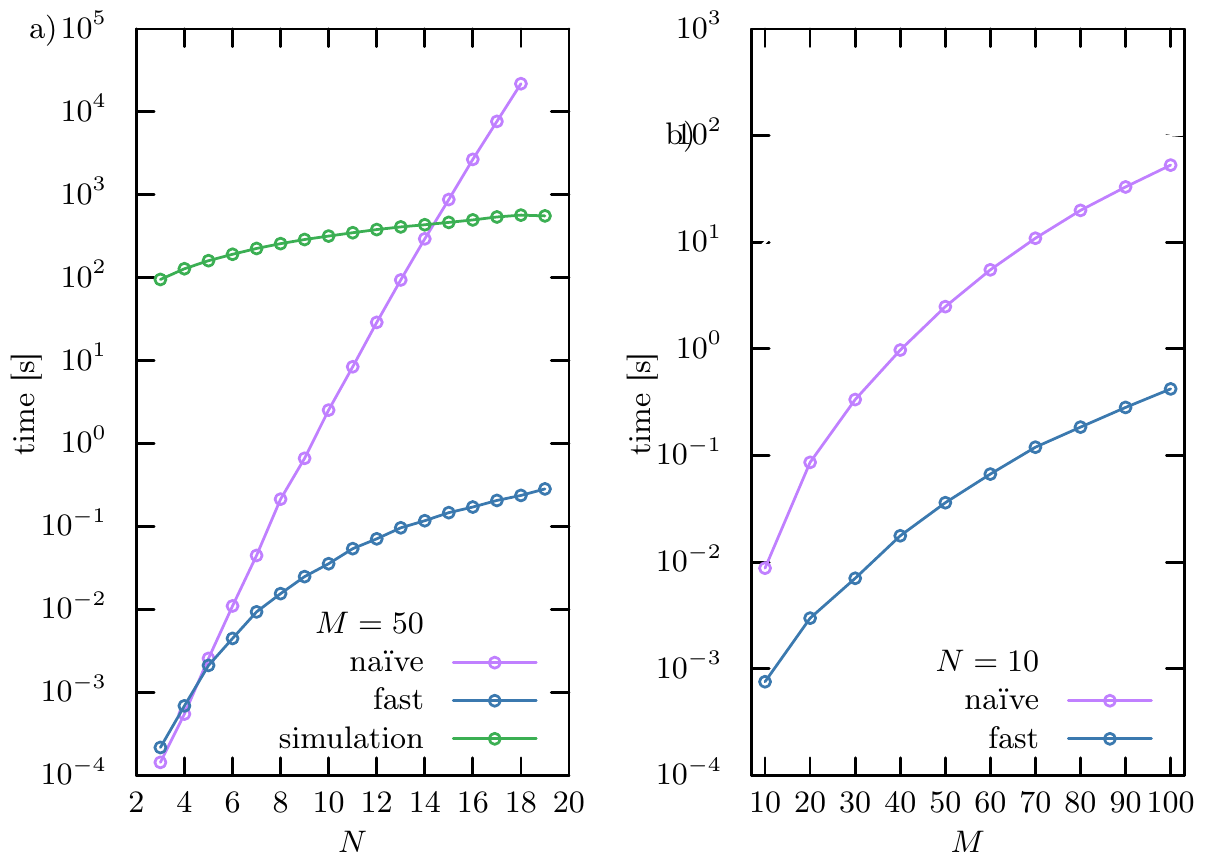}
    \caption{Computational time for calculating the
      {tagged-particle}  Green's function,
      in one specific space-time point a {single-file}
      in a flat potential. We compare our algorithm (blue lines) with
      the na\"{i}ve implementation (purple lines) {to
        a Brownian Dynamics simulation of
        $10^4$ trajectories with $2\times 10^4$ integration-steps
        of length $10^{-3}$ using Algorithm \ref{simulation
          algorithm}}. In a) we fix the max many-body eigenvalue and
      tag the central particle while in b) we fix the total number of
      particles of the {single-file} and tag the fifth
      particle. For both plots the time to calculate all the available
      multisets in Eq.~(\ref{green}) has not been taken into account
      since it is the same  {the na\"ive and efficient implementations of
      the CBA solution}.}   
    \label{running_time}
   \end{figure}

 \paragraph{Exceptions}
 Two classes for managing exceptions:\\
 \codebox{class NotImplementedException : public std::logic_error;}
 and\\
 \codebox{class NotAllowedParameters : public std::logic_error;}
 are included in the code base. The former allows to write a partially implemented derived class, while the latter just throws in the case that an ill-posed parameter is provided. All exceptions throw without any attempt to catch them.
 
 \paragraph{Parallelization}
 By construction the evaluation of Eq.~(\ref{green}) for different $x,\tau$ and
 $x_0$ is parallelizable. A non-trivial 
 parallelization may be achieved implementing a reduction for
 Eq.~(\ref{green}). However, for many systems (see Fig.~\ref{number
   multiset}) the number of terms in the sum is relatively small and
 {a parallel approach is unnecessary unless the
   {single-file} is very big and/or we are interested
   in very small time-scales}. {The present code} does not support parallelization and thread-safety is not guaranteed.
 
 \section{Comparison with Brownian Dynamics simulations}
 \label{comparison simulation}
{A \emph{fair} comparison between our algorithm implementing a
  solution based on Eq.~(\ref{NMGreen}) and a Brownian Dynamics simulation
  integrating the Langevin equation (\ref{Ito}) numerically is somewhat tricky. The reason
  is that the computational effort of a Brownian Dynamics simulation
  grows ``forward'' in time, while the eigenexpansion solution becomes
  challenging ``backward'' in time. In the former case the longer the
  time-scale we are interested in the more integration steps we must
  perform, while if we are interested in shorter time-scales a smaller
  integration time-step must be used. In the implementation of the
  Bethe ansazt  solution in Eq.~(\ref{NMGreen}) we need to consider
  more and more terms in the sum over Bethe eigenvalues
  $\Lambda_{\mathbf{k}}$ in order to obtain reliable results for
  shorter time-scales. In contrast, essentially only two terms
  (i.e. the ground state $\Lambda_{\mathbf{0}}$ and the first
  excited state $\Lambda_{\mathbf{1}}$) are required if we are
  interested only in the long-time dynamics, i.e. $t> 1/\Lambda_{\mathbf{1}}$.}

  {In Algorithm
  \ref{simulation algorithm} we present a convenient method to
  simulate {single-file diffusion based on the Jepsen
    mapping \cite{jepsen_dynamics_1965}}. The key step is the sorting of
  the particles' positions (step 6) that allows avoiding a costly chain
  of \textsf{if} statements required to implement non-crossing
  conditions. To perform this step we use the sorting routine \codebox{std::sort} included in the C++ standard
  library. A comparison of this algorithm with the analytical solution
  can be found in \cite{lapolla_unfolding_2018}.}

 \begin{algorithm}
  \caption{Brownian Dynamics simulation of a single-file}
  \begin{algorithmic}[1]
  \Require \textcolor{white}{.} \begin{itemize}
            \item number of particles $N$;
            \item time-step $\Delta t$, final time $t_f$, list of sampling times $t_s$;
            \item number of trajectories $N_t$;
            \item the initial position of the tagged-particle $x_0$;
            \item a function to update a histogram.
   \end{itemize}
 
   \For {$i$=1:$N_t$}
    \State $t\gets0$;
    \State Generate the initial position for the $N-1$ particle from
    their equilibrium distribution conditioned on the position of the
    tagged-particle $x_0$;
    \While {$t\leq t_f$}
     \State Integrate the Langevin equation (using a Euler-Mayurama scheme for example) of $N$ independent particles using the time-step $\Delta t$;
     \State Sort in ascending order the particles' positions to satisfy the non-crossing condition;
     \If { $t \in t_s$}
      \State Update the histogram containing the Green's function of the tagged-particle;
     \EndIf
     \State $t+=\Delta t$;
    \EndWhile
  \EndFor\\
  \Return the histogram;
  \end{algorithmic}
  \label{simulation algorithm}
 \end{algorithm}
 
  {In Fig.~\ref{running_time} we present the computational time
  required to evaluate the Green's function in a single point in space
  and time  fixing either the maximum eigenvalue (panel a) or the number of
  particles (panel b). In the left panel we also plot the time
  required to compute the tagged-particle Green's function of the single-file in a flat
  potential for different number of particles $N$ by means of a
  Brownian Dynamics simulation. We simulate  $10^4$
  trajectories with a time-step of $10^{-3}$ until time $t_f=20\approx
  2/\Lambda_{\mathbf{1}}$ to ensure that the final equilibrium
  distribution is reached. Using these parameters the statistical error of the simulation  is
  $\sim 5\%-10\%$ using $10^4$ trajectories and $1\%-2\%$ if instead
  we generate $10^5$ trajectories, in agreement with the Gaussian
  central limit theorem. Note that since we are considering enough terms in the series expansion
  (\ref{NMGreen}) the analytic solution may be reliably considered to
  be exact on the time-scale of interest.
  Because the error of the Brownian Dynamics simulation can be reduced
  by increasing the number of independent trajectories, $N_{\rm traj}$, and since
  Algorithm \ref{simulation algorithm} scales linearly with $N_{\rm
    traj}$, it is easy to extrapolate from Fig.~\ref{running_time} the
  computational effort required to obtain more accurate results.}
  
  {Conversely, if we are interested only in short time-scales
  we must carry out the numerical integration of Eq.~(\ref{Ito})  for a small
  number (say $\sim 100$) of steps and thereby obtain better results
  (and with less computational effort) than the analytic
  solution.  This is so because the analytical solution suffers from
  the Runge-phenomenon for short times, since we are approaching a
  delta-function distribution. However, such very short time-scales
  are less interesting since the tagged-particle behaves like a
  free-particle for times shorter the the average collision-time with
  neighboring particles \cite{lizana_single-file_2008}. 
  For the same reason a smaller integration time-step must also be taken to
  capture all the non-trivial physics in a Brownian Dynamics
  simulation if
  we consider a single-file with a large number of particles $N$.
In addition, if the Green's function of the tagged-particle is peaked, 
the binning of the histogram in the analysis of simulations must be
made sufficiently small, which imposes additional constraints on the
integration time-step in order to obtain reliable results.}

 \section{Conclusions}
  We presented an efficient numerical implementation of the exact
 coordinate Bethe ansatz solution of the
 {non-Markovian tagged-particle} propagator in
 a {single-file} in a general confining potential. Motivated by the fact
 that the Bethe eigenspectrum solution nominally carries a
 {large computational cost when the number of
   particles is large} we developed an efficient algorithm, which
 enables investigations of {tagged-particle} diffusion
 on a broad span of time-scales and for various numbers of
 particles. {Our code exploits exchange symmetries in
   order to reduce the number of combinatorial operations.}
 One of the main advantages of the Bethe ansatz
 solution, aside from the fact that it provides an exact solution of
 the problem {and that it ties the tagged-particle dynamics to
many-body relaxation eigenmodes},  is that it is easy to generalize {to take into account for} any confining
 external potential or initial condition. For this reason we provide a
 header file \emph{SingleFileBluePrint.hpp} that allows an easy
 extension of our codebase. With this goal in mind the
 expressiveness and the tools of modern C++ were used to achieve modularity and
 simplicity of use. The code can be easily extended in order to allow
 for a calculation of
 other key quantities related to the non-Markovian dynamics of a
 {tagged-particle}, e.g. the mean square displacement
 \cite{lizana_diffusion_2009} as well as local-time statistics and
 other local additive functionals of {tagged-particle} trajectories \cite{ lapolla_unfolding_2018,lapolla_manifestations_2019}.
 
 \section*{Acknowledgements}
 The financial support from the German Research Foundation (DFG) through the Emmy Noether Program GO 2762/1-1 (to AG), and an IMPRS fellowship of the Max Planck Society (to AL) are gratefully acknowledged.
\appendix

\section{Combinantorics }
\label{permutations}
\paragraph{Permutations}
Let $\mathbf{k}$ be a multiset of $N$ elements. Let $r_i$ denote the multiplicity of each of the $m$ distinct elements of $\mathbf{k}$ such that $\sum_{i=1}^m r_i=N$. Then the number of \emph{distinct permutations} of this multiset is
\begin{equation}
  \binom{N}{r_1,r_2,\cdots ,r_m}=\frac{N!}{\prod_{i=1}^m r_i!}.
  \label{PM}
\end{equation}
The denominator of Eq.~(\ref{PM}) is what we call the multiplicity
$m_\mathbf{k}$ of the multiset $\mathbf{k}$. For example, the distinct
permutations of $\{1,1,2,3\}$ are:
$\{\{1,1,2,3\}$,
$\{1,1,3,2\}$,
$\{1,2,1,3\}$,
$\{1,2,3,1\}$,
$\{1,3,1,2\}$,
$\{1,3,2,1\}$,
$\{2,1,1,3\}$,
$\{2,1,3,1\}$,
$\{2,3,1,1\}$,
$\{3,1,1,2\}$,
$\{3,1,2,1\}$,
$\{3,2,1,1\}\}$.
\paragraph{$t$-combinations}
The problems of enumerating and computing the combinations of a multiset can be mapped to the equivalent bounded composition problems \cite{knuth_donald_ervin_art_2013}. James Bernoulli in 1713 enumerated them for the first time, observing that the number of the $t$-combinations of a multiset $\mathbf{k}$ with $m$ distinct elements, where each of them is contained $r_m$ times, is equal to the $t$-th coefficient of the polynomial $P_\mathbf{k}(z)$:
\begin{eqnarray}
 &P_\mathbf{k}(z)=\prod_{i=1}^m p_i(z)\\
 &p_i(z)=\sum_{j=0}^{r_i} z^j.
\end{eqnarray}
For example the $2$-combinations of $\{1,1,1,2,3\}$ are:
  $\{\{1,1\}$,
  $\{1,2\}$,
  $\{1,3\}$,
  $\{2,3\}\}$.
{\paragraph{Integer partitions}
  We refer to the integer partition of a number $N$ in $m$ parts as the number of ways in which $N$ can be expressed as a sum of all the numbers smaller or equal to itself. For example for $N=4$ and $m=4$:
  $\{0,0,0,4\}$,$\{0,0,1,3\}$,$\{0,0,2,2\}$,$\{0,1,1,2\}$,$\{1,1,1,1\}$.}
\section{Single-particle {eigenspectra}}
\label{single eigenexpansion}
 The dynamics of a single Brownian particle in a unit box in a
 constant potential with reflecting external boundary conditions is
 governed by the Sturm-Liouville problem
 \begin{eqnarray}
   &(\partial_\tau- D\partial^2_x )\Gamma(x,\tau|x_0)=0,&\nonumber\\
   &\partial_x \Gamma|_{x=0}=\partial_x \Gamma|_{x=1}=0&
 \end{eqnarray}
 with the initial condition $\Gamma(x,0|x_0)=\delta(x-x_0)$.
  The corresponding Green's function can be expressed in terms of a spectral expansion
  \begin{equation}
   \Gamma(x,\tau|x_0)=\sum_k \psi^R_k(x)\psi^L_k(x_0)\mathrm{e}^{-\lambda_k \tau},
  \end{equation}
  where 
  \begin{eqnarray}
   &\psi^L_0(x)=\psi^R_0(x)=1,\\
   &\psi^L_k(x)=\psi^R_k(x)=\sqrt{2}\cos(k\pi x),\\
   &\lambda_k = k^2 \pi^2.
  \end{eqnarray}
  On the other hand, if we add a linear potential the corresponding
  Fokker-Planck equation for the Green's function becomes
  \begin{equation}
   (\partial_\tau -D\partial^2_x -g\partial_x) \Gamma(x,\tau|x_0)=0,
  \end{equation}
  with initial condition $\Gamma(x,0|x_0)=\delta(x-x_0)$,
  and the eigenexpansion is given by
  \begin{eqnarray}
   &\lambda_0=0,\\
   &\lambda_k =D k^2 \pi^2+\frac{g^2}{4D},\\
   &\psi^L_0(x)=1,\\
   &\psi^L_k(x)=\displaystyle{\frac{\mathrm{e}^{\frac{gx}{2D}}}{\sqrt{1/2+2Dk^2\pi^2/g^2}}}(\sin(k\pi x)-2Dk\pi\cos(k\pi x)/g),\\
   &\psi^R_0(x)=\displaystyle{\frac{g}{D}\frac{\mathrm{e}^{-\frac{gx}{D}}}{1-\mathrm{e}^{-\frac{g}{D}}}},\\
   &\psi^R_k(x)=\displaystyle{\frac{\mathrm{e}^{-\frac{gx}{2D}}}{\sqrt{1/2+2Dk^2\pi^2/g^2}}}(\sin(k\pi x)-2Dk\pi\cos(k\pi x)/g).\\
  \end{eqnarray}
  Finally for an Ornstein-Uhlenbeck process with natural boundary
  conditions (i.e. $\lim_{|x|\to\infty}\Gamma(x,\tau|x_0)=0$) the Green's
  function is given by
  \begin{equation}
   (\partial_\tau - D\partial^2_x -\gamma \partial_xx) \Gamma(x,\tau|x_0)=0,
  \end{equation}
  with initial condition $\Gamma(x,0|x_0)=\delta(x-x_0)$
  and eigenexpansion
  \begin{eqnarray}
   &\psi^L_k(x)=\displaystyle{\frac{1}{\sqrt{2^k k!}}}H_k(x\sqrt{\gamma/2D})\\
   &\psi^R_k(x)=\displaystyle{\sqrt{\frac{\gamma}{2\pi D}}}\mathrm{e}^{-\frac{\gamma x^2}{2D}}\psi^L_k(x)\\
   &\lambda_k=\gamma k,
  \end{eqnarray}
 where $H_k(x)$ denotes the $k$th ``physicist's'' Hermite polynomial \cite{abramowitz_milton_and_stegun_irene_a_handbook_1964}.


\pagebreak
\bibliographystyle{elsarticle-num}
\bibliography{bethe_article}







\end{document}